\documentclass{PoS}
\usepackage{amsbsy,amssymb,latexsym,amsfonts,amsmath}
\usepackage{graphicx,color}

\newcommand{\A}{\textrm{a}\,}

\newcommand{\bmp}{\noindent\begin{minipage}{16cm}}

\title{The Gradient Flow Coupling in Minimal Walking Technicolor}

\ShortTitle{The Gradient Flow Coupling in Minimal Walking Technicolor}

\author{\speaker{Jarno Rantaharju}
         \thanks{A footnote may follow.}\\
        RIKEN AICS\\
        E-mail: \email{jarno.rantaharju@riken.jp}}

\abstract{We present a measurement of the running coupling in SU(2) with two adjoint fermions in the Yang-Mills gradient flow scheme. The simulations are performed with Schr\"odinger Functional boundary conditions using an improved HEX-smeared Wilson fermion action. We obtain a step scaling function by defining the coupling at a scale relative to the finite size of the lattice. We find a continuum limit with a non-trivial infrared fixed point.}

\FullConference{31st International Symposium on Lattice Field Theory LATTICE 2013\\
                 July 29 August 3, 2013\\
                 Mainz, Germany}

\begin{document}

\section{Introduction}

In the space of gauge field theories coupled to fermions, there is a range in the number of colors and fermion flavors where the $\beta$-function has a non-trivial zero in the infrared.
These conformal theories have applications in model building beyond the standard model, particularly in technicolor theories, where the electroweak symmetry is broken by a spontaneous symmetry breaking in a strongly interacting model.
They are also interesting from the theoretical point of view of understanding the structure of gauge field theories.

When the number of fermion flavors is large, close to the upper limit of the conformal window, the coupling at the fixed point is small and perturbation theory is expected to apply.
When approaching the lower limit, however, the fixed point coupling grows and non-perturbative methods are needed.
The $\beta$-function can be studied in lattice field theory by studying the behavior of a renormalized coupling when the renormalization scale is changed \cite{Luscher:1991wu}.

Close to the fixed point the coupling runs very slowly and high accuracy is required to discern the direction of the running and to find the fixed point.
To improve the accuracy, we study the gradient flow method for measuring the coupling.
The method has been studied previously in several models 
and with varying boundary conditions
\cite{Fodor:2012qh,Fritzsch:2013je,Fritzsch:2013hda,Ramos:2013gda}.
Compared to the more standard Sch\"odinger Functional method,
the statistical accuracy of the gradient flow method is higher.
However, in the form presented here, the method also has larger discretization errors.

We study the coupling in
the SU(2) gauge theory coupled with two fermions in the adjoint representation.
In previous studies using the Schr\"odinger functional method the model
has exhibited a non-trivial infrared fixed point at a relatively
small coupling
\cite{Catterall:2007yx,Hietanen:2008mr,
    DelDebbio:2008zf,Catterall:2008qk,Hietanen:2009az,
    Bursa:2009we,DelDebbio:2009fd,DelDebbio:2010hx,
    DelDebbio:2010hu,Bursa:2011ru,DeGrand:2011qd,
    Giedt:2012rj,Rantaharju:2013gz}.
It is also relatively fast to simulate due to its minimal gauge and fermion content,
making it a good test case for the gradient flow method.

\section{Gradient Flow Coupling}

Several applications of the gradient flow in lattice simulations have been studied recently
\cite{Luscher:2009eq,Luscher:2010iy,Deuzeman:2012jw,Luscher:2013cpa,Suzuki:2013gza,DelDebbio:2013zaa}.
In gradient flow, the gauge field is allowed to develop in
a fictitious flow time along the gradient of an action.
The flow defined by the continuum gauge action is given by
\begin{align*}
  \partial_t B_{t,\mu} &= D_{t,\mu} B_{t,\mu\nu}, \\ B_{0,\mu} &= A_\mu\\
  B_{t,\mu\nu} &= \partial_\mu B_{t,\nu} - \partial_\nu B_{t,\mu} 
  + \left[ B_{t,\mu},B_{t,\nu} \right].
\end{align*}
Here $B_{t,\mu}$ is the flow field parametrized by the flow time $t$, and $A_\mu$ is the original gauge field.
The flow naturally smooths the field, moving toward the minimum of the action.
Remarkably, the flow field $B_{t,\mu}$, at any finite flow time, is a finite renormalized field.
The correlators of the flow field are automatically renormalized
and therefore encode physical properties of the theory \cite{Luscher:2011bx}.

As a representative case, the observable 
\begin{align*}
  \left<E(t)\right> &= \frac 14 \left<G_{\mu\nu}(t)G_{\mu\nu}(t)\right>
\end{align*}
was studied in \cite{Luscher:2010iy}.
To the leading order in perturbation theory, it has the form
\begin{align*}
  \left<E(t)\right> &= N g^2/t^2 + \mathcal{O}(g^4).
\end{align*}
The observable can therefore be used to define a renormalized coupling, \cite{Fodor:2012qh}
\begin{align*}
  &g^2_{GF} = \frac{t^2 \left < E(t) \right>}{N}.
\end{align*}

We study the running of the coupling with respect to the renormalization scale
by measuring the coupling with several physical lattice sizes.
To quantify the change we use the step scaling function \cite{Luscher:1993gh}
\begin{align}\label{eq:sigma1}
  &\Sigma(u,\A/L) = \left . g_{GF}^2(g_0,2L/\A) \right|_{g_{GF}^2(g_0,L/\A)=u}\\\nonumber
  &\sigma(u) = \lim_{\A \rightarrow 0} \Sigma(u,\A/L). \nonumber
\end{align}
The step scaling function describes how the coupling evolves when
the linear size of the system changes from $L$ to $2L$.
The gradient flow introduces another length scale, $l=\sqrt{8t}$,
which we also need to change
when varying the lattice size.
In practice we choose $l=0.5L$.

We use the same highly improved action as in our
Sch\"odinger functional calculations to allow direct
comparison of the results.
Preliminary results from this study
have already been published \cite{Rantaharju:2013gz}
and the full results will be published soon.
The action is given by
\begin{align*}
  S = (1-c_g)S_G(U) + c_g S_G(V) + S_F(V) + c_{SW} \delta S_{SW}(V),
\end{align*}
where $S_G(U)$ is the standard Wilson gauge action
and $S_G(V)$ is the gauge action calculated using HEX smeared links \cite{Capitani:2006ni}.
The gauge smearing parameter $c_g$ is set to $0.5$.
$S_F(V)$ and $\delta S_{SW}(V)$ are the Wilson fermion action
and the clover correction using HEX smeared links.
The clover coefficient takes the tree level value $c_{SW}=1$.

We perform the simulations using Schr\"odinger functional boundary conditions.
They have some advantages over fully periodic boundary conditions:
The gauge field has a unique global minimum, which simplifies perturbation theory
and removes non-even orders of coupling from the perturbative expansion of
the $\beta$-function \cite{Fritzsch:2013je}.
The boundary conditions also alleviate problems with massless fermions.
For the gauge fields they are given by
\begin{align*}
  U_k(x) = 1 \text{, when } x_0=0,L \\
  U_\mu(x+L\hat k) = U_\mu(x).
\end{align*}
The fermion fields are set to zero at the time boundaries
and have twisted periodic boundary condition in the spacial directions:
\begin{align*}
  \psi(x + L\hat k) = e^{i\pi/5}\psi(x).
\end{align*}

Since the translational symmetry is now broken in the time direction,
we avoid any unnecessary discretization errors by
only using the middle time slice for the measurement,
\begin{align}
   g^2_{GF} = \frac{t^2 \left < E(t,x_0) \right>}{N}, \,\,\,\,\, x_0=L/2, \,\, t=c^2L^2/8.
   \label{eq:gfcoupling}
\end{align}
Because of the improved lattice action, the normalization factor $N$ is unknown on the lattice.
We use the continuum value $N = 288.527$.

\section{Results}

\begin{figure}
  \includegraphics[width=0.45\textwidth]{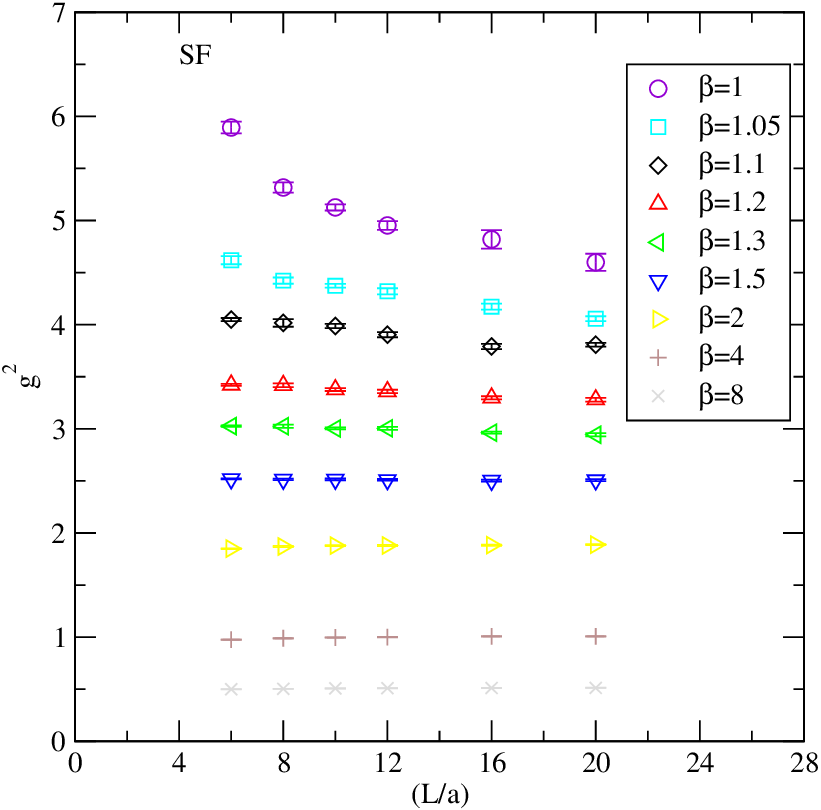} 
  \includegraphics[width=0.45\textwidth]{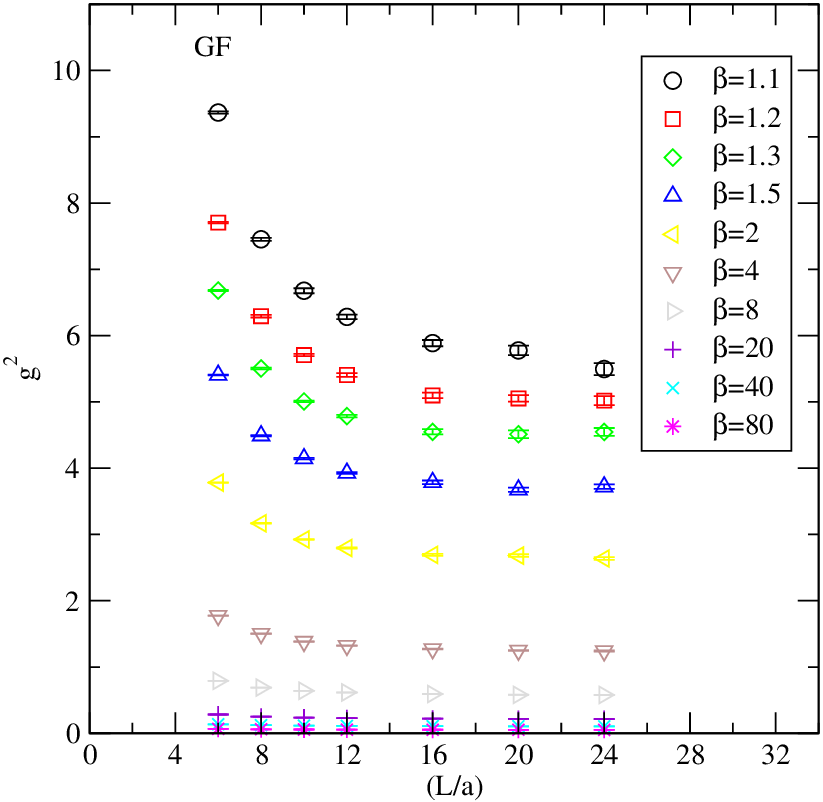}
\caption[b]{
  The measured values of the coupling using the Schr\"odinger functional
  and the gradient flow methods.
  The plot on the left shows the SF coupling and the plot on the right shows the GF coupling.
}
\label{fig:coupling}
\end{figure}

\begin{table}
\centering
\begin{tabular}{|c|c|c|c|c|c|} 
\hline
  \, & \multicolumn{2}{|c|}{ SF } & \multicolumn{3}{|c|}{ GF } \\
  \hline
  $L/\A$ & 20 & 20 & 20 & 20 &24 \\
  $\beta=4/g_0^2$ & 2 & 1.2 & 2 & 1.2 & 1.2 \\
  $N_{\operatorname{traj}}$ & 151203 & 158828 & 14065 & 13693 & 9648 \\
  $(\delta g^2/g^2)_{\operatorname{statistical}}$ & 0.0059 & 0.0109 & 0.0112 & 0.0121 & 0.0150 \\
  $\left(\delta g^2/g^2\right)^2 N_{\operatorname{traj}}$ &
  5.225 & 18.75 & 1.765 & 2.048 & 2.250 \\
  $t_{\operatorname{traj}}$ & 1221.11 & 1277.04 & 585.58 & 769.64 & 1787.13 \\
  \hline
\end{tabular}
\caption[b]{
  The statistical errors of the SF and the GF couplings at two values of $\beta$
  with large lattice sizes.
  $\left(\delta g^2/g^2\right)^2 N_{\operatorname{traj}}$ estimates the number of
  trajectories required to reach a given statistical error.
}
\label{table:1}
\end{table}

\begin{table}
\centering
\begin{tabular}{|c|c|c|c|c|c|c|c|} \hline
  $L/\A$ & 6 & 8 & 10 & 12 & 16 & 20 & 24 \\ \hline
  SF, $\beta=2$   & 0.093 & 0.259 & 0.265 & 0.880 & 3.791 & 5.225 &    \\
  GF, $\beta=1.2$ & 0.577 & 0.373 & 0.988 & 0.885 & 1.191 & 2.048 & 2.250 \\
  \hline
\end{tabular}
\caption[b]{
  The volume dependence of the error estimate $\left(\delta g^2/g^2\right)^2 N_{\operatorname{traj}}$ in the SF method and the GF method.
}
\label{table:2}
\end{table}

We compare the gradient flow coupling with the Schr\"odinger functional coupling
measured in the same model.
The Schr\"odinger functional and gradient flow couplings are measured with constant values of the bare coupling. All bare parameters, except the boundary conditions, are identical in the two sets of simulations.

It is clear already from the lattice values given in figure \ref{fig:coupling}, that the gradient flow coupling has large discretization errors. The coupling should depend only little on the scale, especially close to the fixed point, but the gradient flow coupling shows strong dependence on the lattice size. Naturally the figure does not tell us how the discretization errors approach the continuum limit.

We also see that the statistical errors are roughly equal.
The SF measurements are calculated from around 150 000 HMC trajectories of length 2 
and the GF measurements from around 15 000 trajectories of length 1.
The behavior of statistical errors is shown in table \ref{table:1} for two
values of $\beta$ and $L/a$. 
It is clear that the GF coupling requires less statistics for the same accuracy.
The statistical accuracy of the GF coupling also scales better with the lattice size,
as table \ref{table:2} demonstrates.

For a more robust comparison we should take a continuum limit of the measurements.
We achieve this by calculating the step scaling function $\Sigma(u,a/L)$ in equation
\ref{eq:sigma1} and taking a continuum limit keeping the renormalized coupling squared, $u$, constant.
Since the measurements are calculated at constant bare coupling, we use an interpolating function to find the value at a given renormalized coupling.
For the SF and GF couplings respectively, the interpolating functions are given by
\begin{align}
  g_{SF}^2(g_0) = g_0^2 \frac{\prod_{k=1}^3 (1+a_k g_0^2)}{\prod_{k=1}^3 (1+b_k g_0^2)}, \,\,\,\,\,\, g_{GF}^2(g_0) = a_0 g_0^2 \frac{\prod_{k=1}^2 (1+a_k g_0^2)}{\prod_{k=1}^3 (1+b_k g_0^2)}. \label{eq:ginter}
\end{align}
Note that at first order, $g^2_{GF}(g_0,L/a) = a_0(L/a) g_0^2$.
Thus, in the continuum limit, $a_0=1$,
and any deviation represents a discretization error
that would have been removed if we had used a lattice result for the normalization constant $N$ in equation \ref{eq:gfcoupling}.
We can therefore define a scaled version of the GF coupling by dividing by $a_0(L/a)$.

\begin{figure}
  \includegraphics[width=0.45\textwidth]{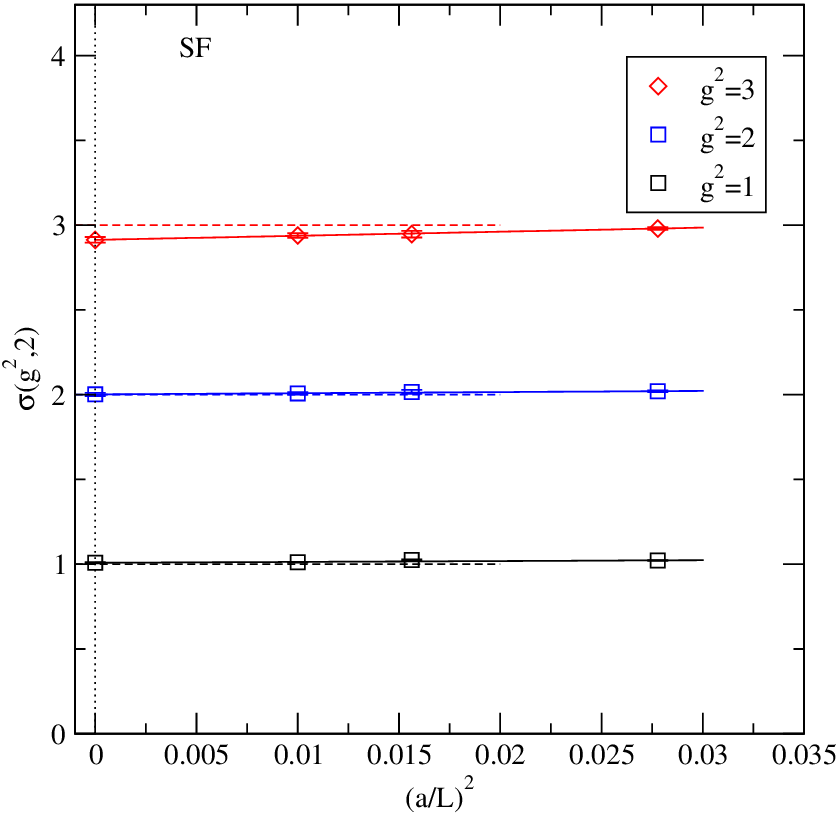} 
  \includegraphics[width=0.45\textwidth]{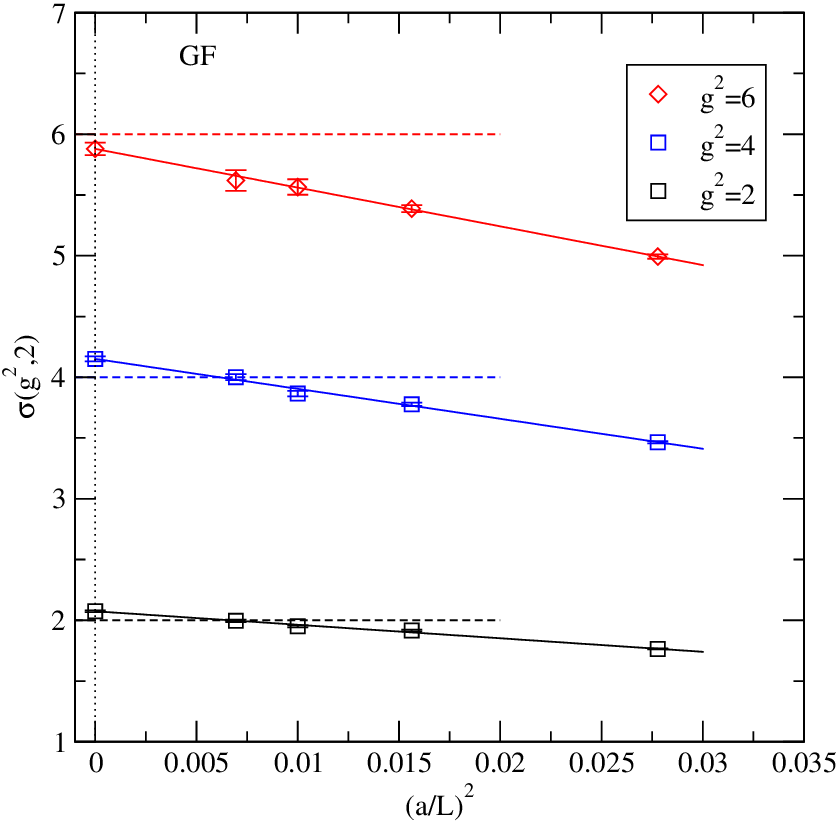}
\caption[b]{
  The continuum extrapolation of the step scaling function.
  The left hand side shows the SF case and
  the right hand side shows the GF case.
}
\label{fig:sigmalim}
\end{figure}

\begin{figure}
  \includegraphics[width=0.45\textwidth]{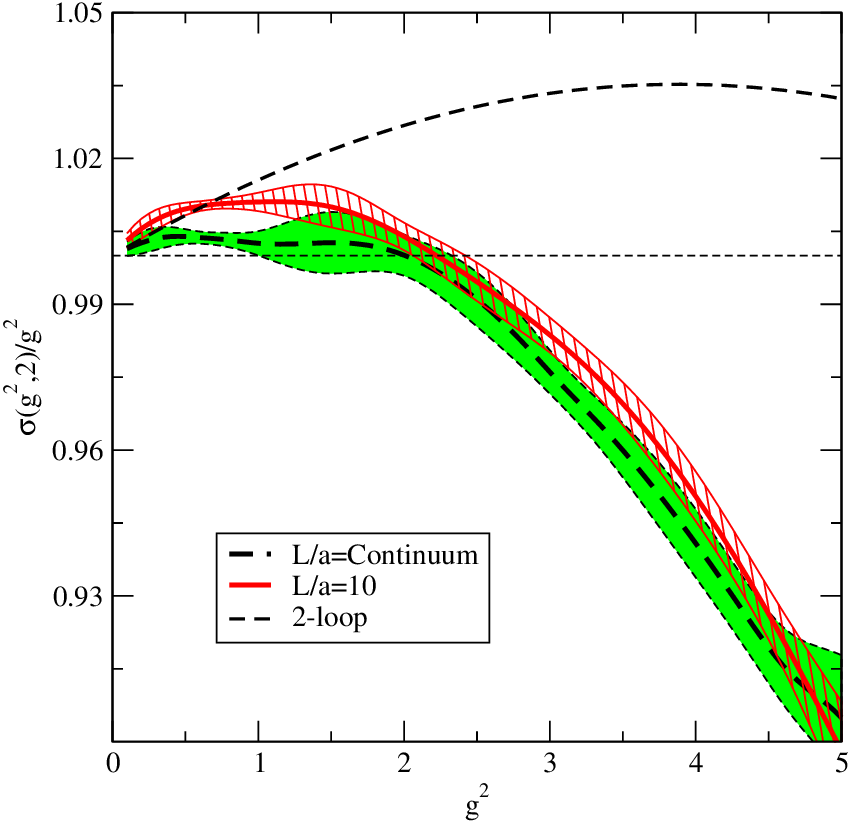} 
  \includegraphics[width=0.45\textwidth]{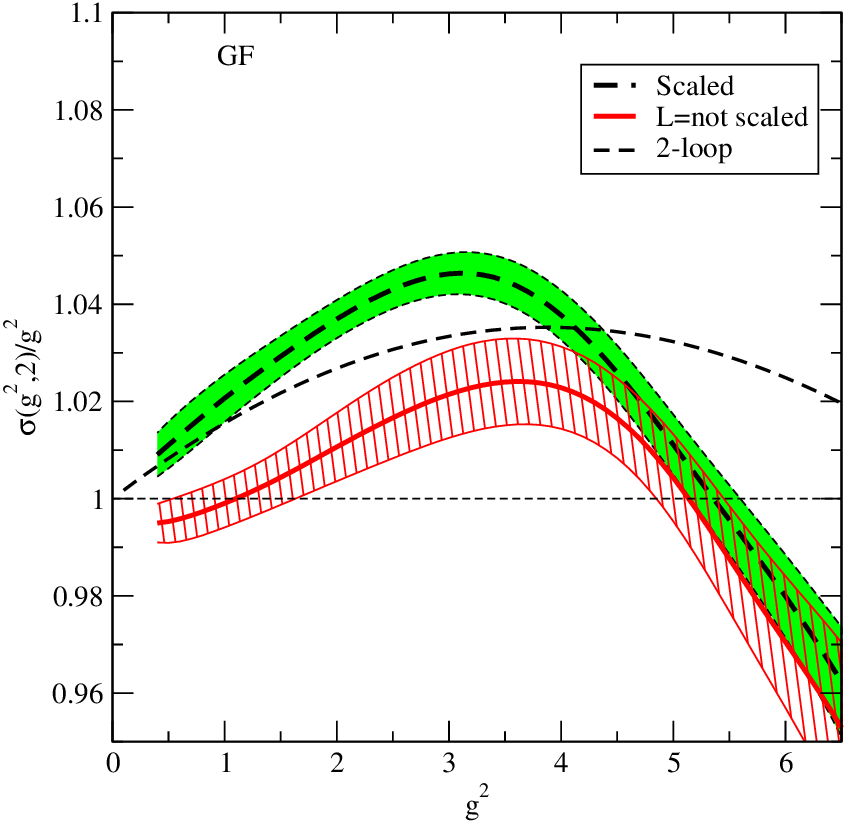}
\caption[b]{
  The scaled continuum step scaling function $\sigma(g^2,2)/g^2$ using the SF and GF methods. 
  On the left:
  The black line with the green error band corresponds to the continuum result.
  The red line with the hashed band correspond to the lattice result at the largest lattice size $L/a=10$.
  The black dashed line gives the 2-loop perturbative result. 
  On the right:
  The black line with the green band correspond the continuum limit with the coupling scaled with $a_0(L/a)$ at each lattice size.
  The red line with the hashed error band corresponds to the continuum result without scaling.
  The black dashed line gives the 2-loop perturbative result. 
}
\label{fig:sigmacont}
\end{figure}

Using the interpolating functions we can now calculate $\Sigma(u,a/L)$ at a constant $u$.
We estimate the continuum limit using a second order extrapolation,
$\Sigma(u,a/L) = \sigma(u) + c (a/L)^2$.
We show the continuum extrapolation for three representative values of $u$
in figure \ref{fig:sigmalim} and the continuum value $\sigma(u)$ in figure \ref{fig:sigmacont}.

The continuum extrapolation is steeper in the GF measurement, but nevertheless consistent with the second order extrapolation.
Both measurements show a fixed point, but the GF method gives a larger coupling at the fixed point.
This is partially due to the difference in the two schemes, but also suggests that there may be higher order discretization effects influencing the results.

\section{Conclusions}

We have studied the gradient flow coupling in the SU2 gauge field theory with two fermions in the adjoint representation with Schr\"odinger functional boundary conditions.
We find that the measurement is in qualitative agreement with the previous studies
using the Schr\"odinger functional method.
The results are promising, as the statistical errors are much smaller than in the Schr\"odinger functional coupling and scale better with the lattice size.
The increased discretization effects present a problem, which may be alleviated using
improved flow equations or definitions of the measurable $\left < E(t) \right>$.
It is also worth studying the effects of the boundary conditions on the coupling.

\end{document}